\documentclass[conference,letterpaper]{IEEEtran}
\usepackage[top=0.75in, bottom=1.05in, left=0.625in, right=0.625in]{geometry}
\usepackage[bookmarks=false, pdfborder={0 0 0}, colorlinks=false]{hyperref}
\usepackage{cite}
\usepackage{amsmath,amssymb,amsfonts}
\usepackage{algorithmic}
\usepackage{graphicx}
\usepackage{caption}
\usepackage{subcaption}
\usepackage{stfloats}
\usepackage{textcomp}
\usepackage{xcolor}
\usepackage{orcidlink}
\def\BibTeX{{\rm B\kern-.05em{\sc i\kern-.025em b}\kern-.08em
    T\kern-.1667em\lower.7ex\hbox{E}\kern-.125emX}}
\begin{document}

\title{Learning from Disengagements: An Analysis of Safety Driver Interventions during Remote Driving\\
\thanks{This study was conducted in collaboration with Vay Technology, a company operating a remotely driven commercial car-sharing service. The analysis focuses on data collected from operations in Las Vegas, Nevada, US, while Vay Technology also conducts remote driving operations in Hamburg and Berlin.}
}

\author{\IEEEauthorblockN{1\textsuperscript{st} Ole Hans\orcidlink{0009-0009-0239-9469}}
\IEEEauthorblockA{\textit{Department of Operational Safety}\\
Vay Technology GmbH\\
Berlin, 12099 Germany \\
ole.hans@vay.io}
\and
\IEEEauthorblockN{2\textsuperscript{nd} Jürgen Adamy\orcidlink{0000-0001-5612-4932}}
\IEEEauthorblockA{\textit{Department of Control Methods and Intelligent Systems}\\
Technical University of Darmstadt \\
Darmstadt, 64289 Germany \\
juergen.adamy@tu-darmstadt.de}
}
\maketitle

\begin{abstract}
This study investigates disengagements of Remote Driving Systems (RDS) based on interventions by an in-vehicle Safety Drivers (SD) in real-world Operational Design Domains (ODD) with a focus on Remote Driver (RD) performance during their driving training. Based on an analysis of over \mbox{14,000 km} on remote driving data, the relationship between the driving experience of 25 RD and the frequency of disengagements is systematically investigated. The results show that the number of SD interventions decreases significantly within the first \mbox{400 km} of driving experience, which illustrates a clear learning curve of the RD. In addition, the most common causes for \mbox{183 disengagements} analyzed are identified and categorized, whereby four main scenarios for SD interventions were identified and illustrated. The results emphasize the need for experience-based and targeted training programs aimed at developing basic driving skills early on, thereby increasing the safety, controllability and efficiency of RDS, especially in complex urban environment ODDs.
\end{abstract}


\section{Introduction}
The remote operation of vehicles is becoming increasingly important, especially in the context of modern mobility concepts and the further development of Automated Driving Systems (ADS). Studies such as the remote driving report published by McKinsey \cite{mckinsey2024remote_driving} and progress of regulations for remote driving in Germany \cite{StVFernLV2024} underline the relevance of the remote driving technology for the future of mobility. 

However, in order to safely introduce this technology to the market, a well-founded examination of central research questions is required, as emphasized in the Federal Highway Research Institute (BASt) research report for teleoperation needs \cite{TeleoperationReport}.

One of these central questions concerns the requirements for the education and training of \mbox{Remote Drivers (RD)}. Existing standards and guidelines, such as the BSI Flex 1887 \mbox{standard \cite{BSIFlex1887}}, demand comprehensive training that focuses on familiarity with different vehicle conditions and realistic training environments. Likewise, the importance of targeted training to improve human performance and mitigate the inherent challenges of remote driving is emphasized in related research work by several authors \cite{Schwindt-Drews2024, hans2025evaluationremotedriverperformance}. 

However, before specific requirements for the education and training of RDs can be formulated, it is crucial to identify the underlying problems. Previous research has often relied on simulations \cite{neumeier2019teleoperation} or controlled test environments \cite{tener2022driving}, often with untrained participants. These approaches provide valuable insights into human behavior under different conditions, but cannot fully reflect the complexity of real-world driving situations \cite{hussain2019speed}. 

The gap between industrial applications and scientific validation is also highlighted by the increasing number of implementations in an industrial context, for example by companies such as Vay Technology or Fernride. While these companies demonstrate the practical potential of remote driving, scientific research into human performance in real-world Operational Design Domains (ODD) remains insufficient.

This work addresses the gap in the scientific investigation of Remote Driving Systems (RDS) on public in real-world ODD and analyses disengagements of the RDS by in-vehicle Safety Driver (SD) interventions. These are situations in which the vehicle control has to be taken over by the in-vehicle SD to prevent a potential hazardous situations. The aim is to use this data to draw conclusions about the requirements for training, system safety design and the specification of ODDs in order to create the basis for a safe and effective market introduction of remote driving.

Section \ref{relatedwork} provides literature research relevant to the study, and Section \ref{SYSTEMMODEL} describes the RDS in its parts and the respective ODD as the basis for this study. Section \ref{AnalysisDataPreparation} describes the data set which is used. In Section \ref{evaluation} the disengagement classification is defined and the disengagement distribution in addition to the impact of  RD experience are evaluated. Based on that Section \ref{analysis_dis_reasons} presents the analysis results of the disengagement reasons in detail. The limitations of the results are described in Section \ref{Limitation}. Finally, the conclusion and proposed further work are presented in Section \ref{conclusion}.

\section{Related Work}
\label{relatedwork}
This Section presents an overview of research related to remote driving, their limitations, and SD disengagements for remote driving.

\subsection{Remote Driving}
\label{remotedriving}
Remote driving in the automotive industry is distinguished from related concepts such as remote assistance and remote monitoring, forming a nuanced taxonomy of Remote Human Input Systems that varies in complexity \cite{bogdoll2022taxonomy}. Remote driving involves the direct control \cite{majstorovic2022survey} of a vehicle by a RD, from a remote location, enabling the navigation of a vehicle through complex environments without requiring physical presence in the vehicle. Building on foundational work by \mbox{Bogdoll et al. \cite{bogdoll2022taxonomy}} and Amador et al. \cite{amador2022survey}, this field is classified by levels of complexity, which aligns with the SAE taxonomy for ADS \cite{SAEI}. The core of remote driving lies in a stable wireless connection between the Remote Control \mbox{Station (RCS)} and the vehicle, enabling real-time monitoring and control. This connection allows the RD to access the vehicle within ODD remotely, using data transmitted from cameras and further sensors, albeit with latency that impacts real-time the RD's situational awareness \cite{kettwich2021teleoperation, neumeier2019teleoperation}. 

Challenges in RDS operation are both technical and human-centered. Technical challenges such as latency, video quality, and visibility impairments create performance constraints for the RDS itself as well as for the RD, while human factors, such as the lack of haptic feedback, are additional factors that reduce the situational awareness of the RD \cite{chen2007human, hans2024human}. The RD relies mainly on visual feedback, which poses unique limitations compared to conventional driving. Latency and reduced video quality can further impede the RD’s capacity to maintain consistent situational awareness while operating the vehicle \cite{tener2022driving, neumeier2019teleoperation}. Given these limitations, a well-defined ODD must reflect not only system capabilities but also the human driver’s performance threshold to ensure effective RD functionality \cite{hans2023operational}. Addressing these challenges demands both functional adaptations and operational measures, emphasizing the need for both technological enhancements and targeted human performance adaption and training to mitigate the inherent limitations of remote driving \cite{Schwindt-Drews2024, hans2024human}. 

\subsection{Disengagements}
\label{disengagements}
Related studies have investigated disengagements mainly in the context of ADSs, where the in-vehicle SD intervenes when the ADS cannot perform the driving task autonomously \cite{cummings2024identifying}. These studies often focus on the frequency of disengagements, the ODD factors that contribute to such events, and the underlying technical failures. Kalra and Paddock \cite{kalra2016driving} analyzed disengagement reports from tests of ADSs to estimate failure rates and potential safety risks. Similarly, publicly available reports from ADS companies such as Waymo \cite{waymo2019disengagements} have provided valuable insights into the circumstances that necessitate human intervention in autonomous driving.
Disengagements during remote driving, in which a SD overwrites the inputs of the RD, have a comparable significance. For remote driving purposes in this study, disengagements can be initiated by the SD either by pressing the accelerator or brake pedal or by moving the steering wheel.

\begin{figure}
\centerline{\includegraphics[width=2.1in]{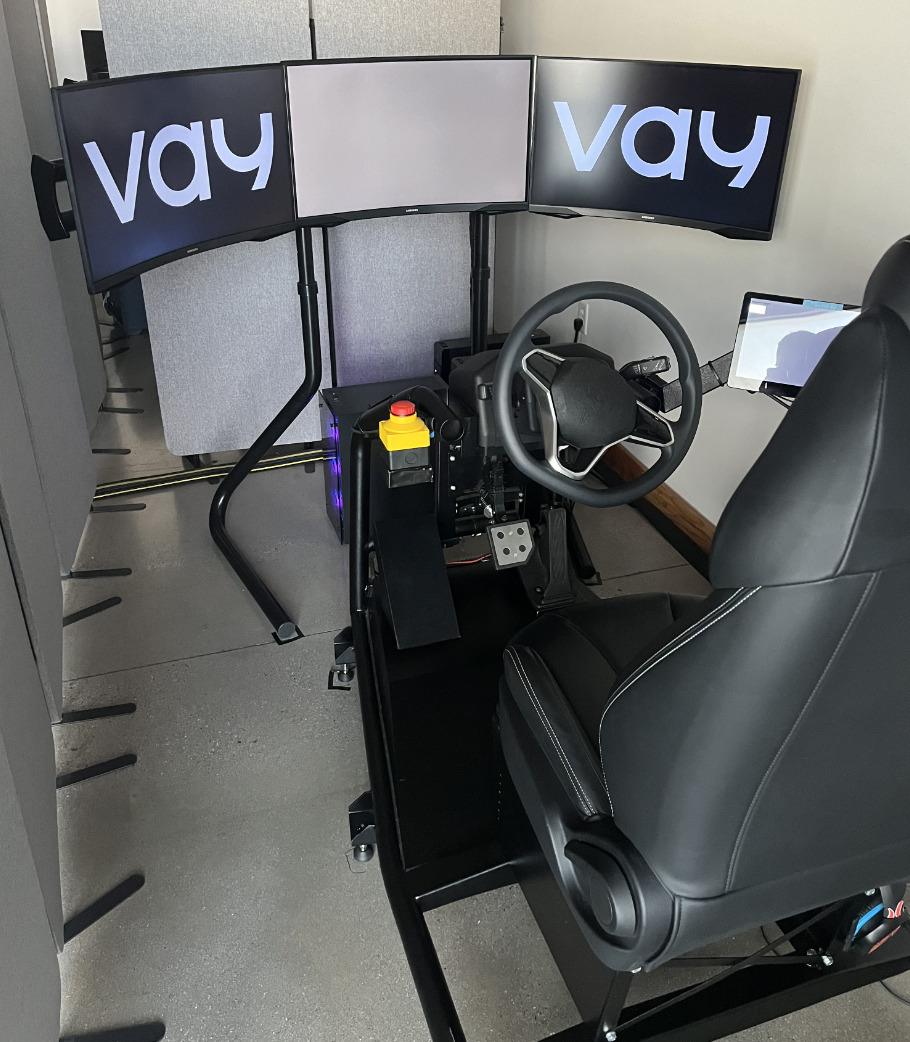}}
\caption{Vay Remote Control Station (RCS) within an operations center in Las Vegas, Nevada.
\label{VayRCS}}
\end{figure}

\section{Remote Driving System Model}
\label{SYSTEMMODEL}
In this work, the RDS developed by Vay Technology was utilized. The RDS consists of three main components: the vehicle, which is equipped with the Vay hardware and software, the RCS, which enables remote control, and the RD as the human-in-the-loop. For the purposes of this work, a SD is used as a additional safety fallback level for RD training purposes. The RD acts as the vehicle driver and communicates any driving intentions or changes in the immediate surroundings. This enables the SD to understand the RD's intended actions. However, if the RD deviates from the announced intention, commits a traffic violation or creates a potentially unsafe situation, the SD takes over control. 

The vehicle was retrofitted to integrate the remote driving technology, including additional cameras and safety controller that monitor and regulate critical safety parameters in real-time. Furthermore, the vehicle’s connectivity was enhanced by specific antennas and modems, integrated with the proprietary Vay connectivity software stack, ensuring a stable and fast communication link between the vehicle and the RCS.

\begin{figure*} 
    \centering
    \begin{subfigure}{0.31\textwidth}
        \includegraphics[width=\linewidth]{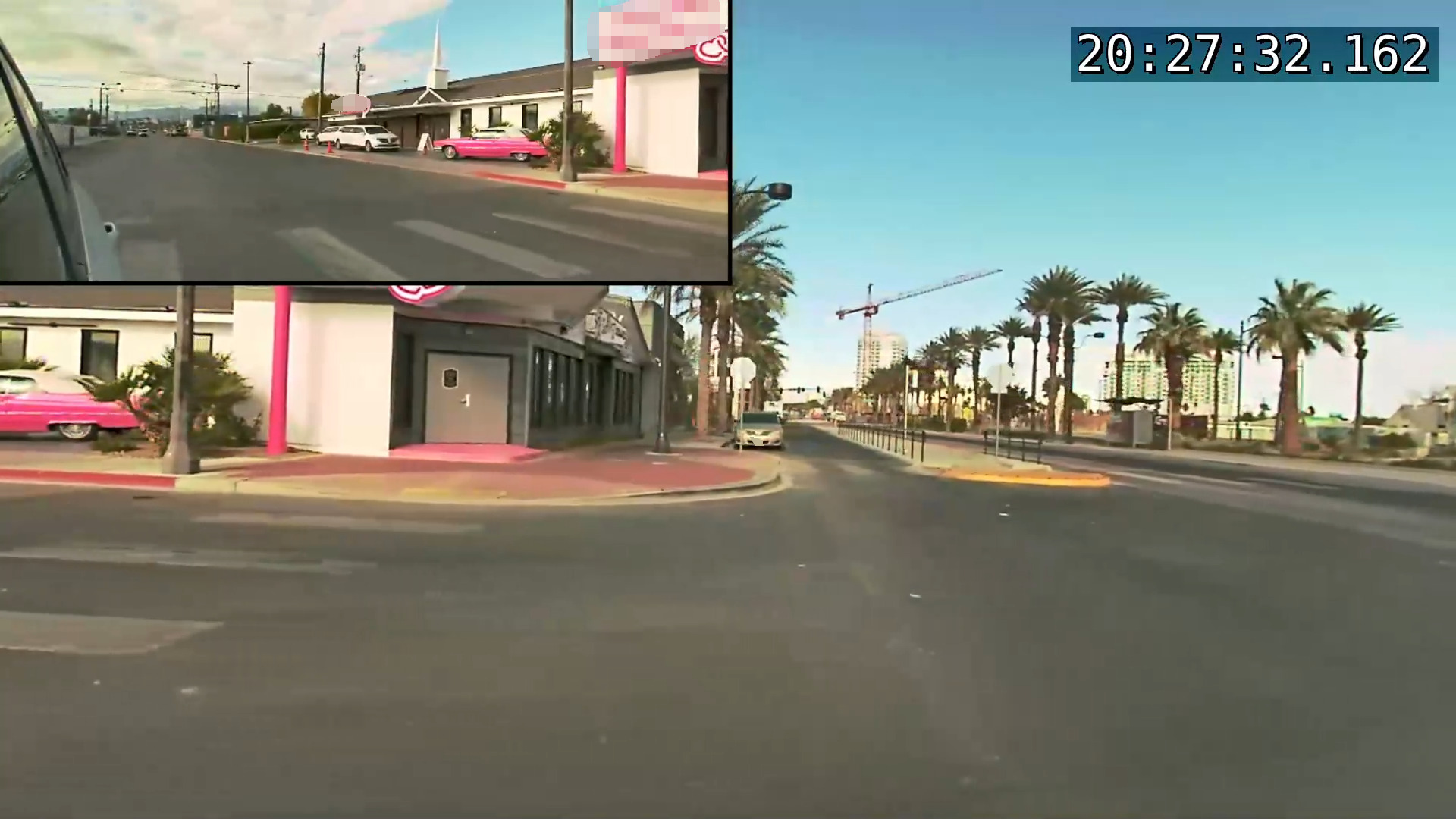}
        \caption{Left Side Screen}
        \label{Left_side_screen}
    \end{subfigure}
    \hfill
    \begin{subfigure}{0.31\textwidth}
        \includegraphics[width=\linewidth]{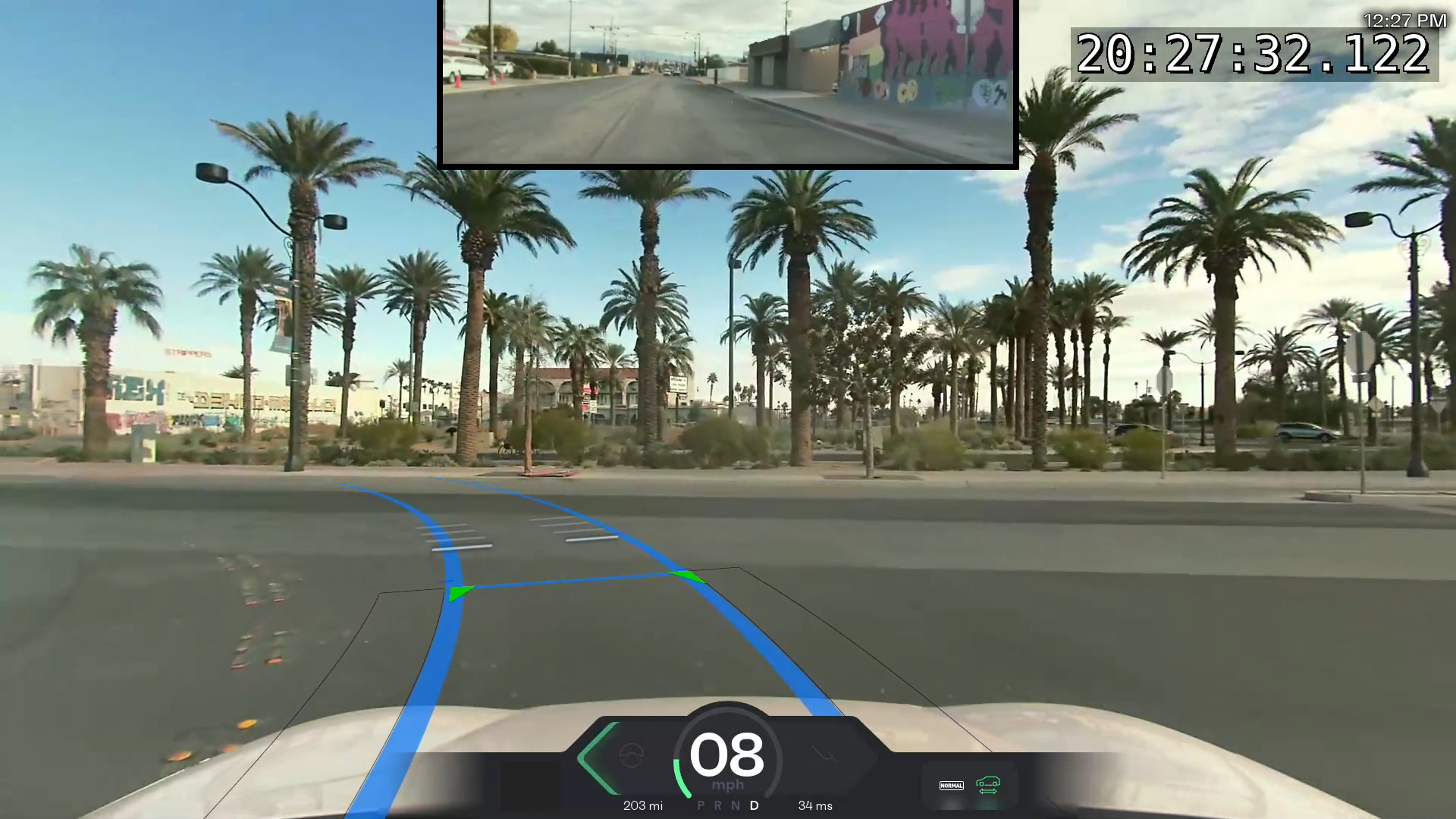}
        \caption{Front Screen}
        \label{Front_screen}
    \end{subfigure}
    \hfill
    \begin{subfigure}{0.31\textwidth}
        \includegraphics[width=\linewidth]{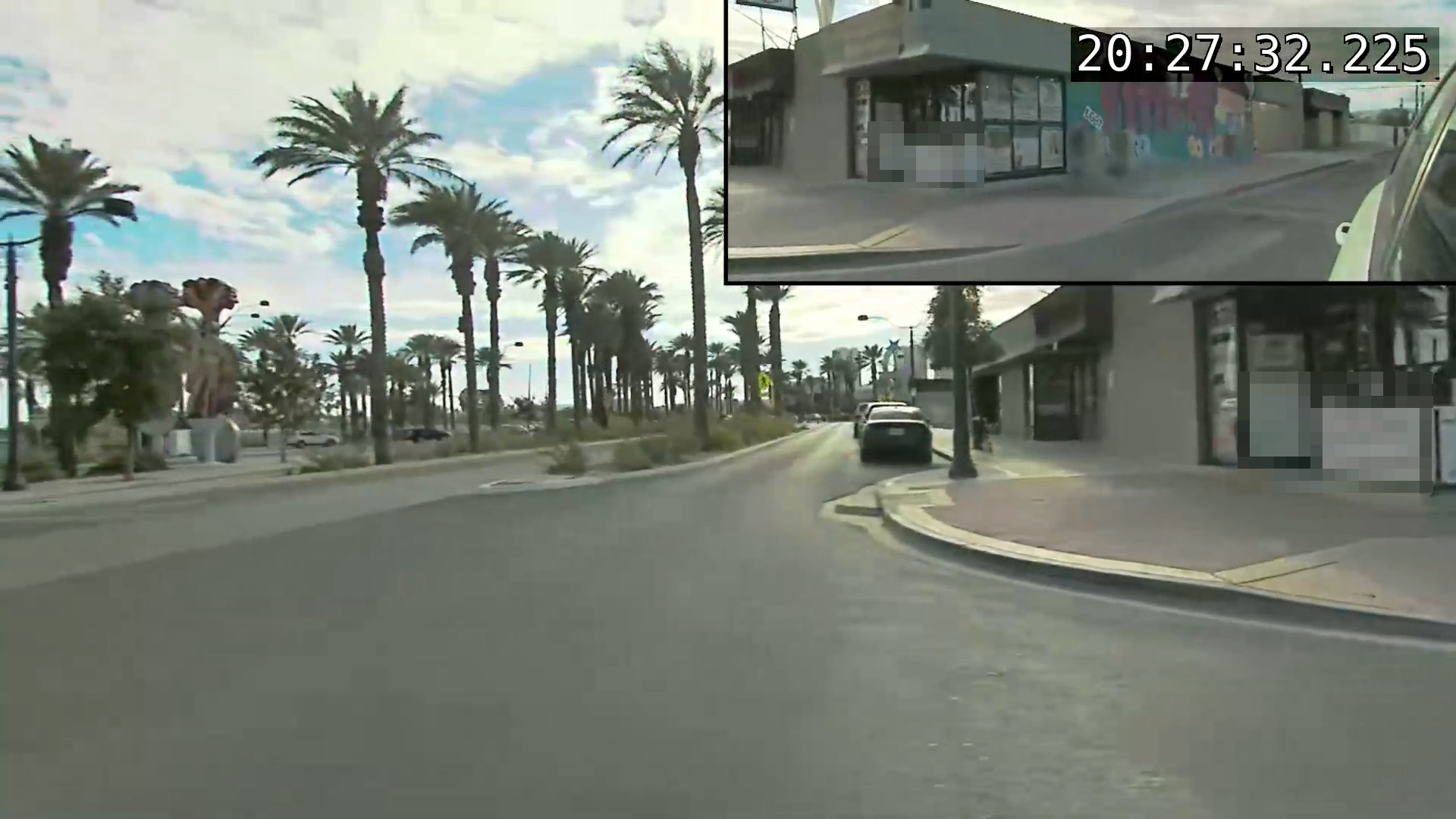}
        \caption{Right Side Screen}
        \label{Right_side_screen}
    \end{subfigure}
    \caption{Human Machine Interface (HMI) of the Remote Control Station (RCS) showing a left turn maneuver.}
    \label{HMI_Interface}
\end{figure*}

The RCS, shown in \mbox{Fig. \ref{VayRCS}}, serves as the Human-Machine-Interface (HMI) for the RD, who operates the vehicle remotely. It features three screens that display visual feedback transmitted through multiple camera sensors from the vehicle's cameras and further sensors, as visualized in \mbox{Fig. \ref{HMI_Interface}}. Additionally, microphones and speakers ensure auditory feedback and communication to the inside of the vehicle, while road traffic sound is delivered via external microphones to the RD's headphone. To make it more realistic for RDs to control a vehicle, a concept of torque emulation is used, introduced by \mbox{Hossain et al. \cite{hossain2024stability}}. In addition to that, steering force feedback is provided to the RD. The vehicle is controlled via a automotive-grade physical steering wheel, along with automotive-grade controls such as column switches, throttle, and brake pedals. Special controllers in the RCS process incoming data and enable interaction with the Vay system installed in the vehicle.

\subsection{Remote Driving System Human-Machine-Interface}
\label{HMI}
As shown in \mbox{Fig. \ref{HMI_Interface}}, the HMI provides the RD with all relevant information on vehicle control and the environment in an intuitive display. It combines visual aids and real-time camera data to enable precise control. The instrument display provides information about the vehicle's status (for example warning lights) and allow for quick identification of technical issues of the base vehicle, as the instrument display in a regular car. In addition, the HMI displays current speed, gear selection, system latency, and remaining range to support efficient and controlled driving. The blue trajectory lines show the planned vehicle path and a so-called safety corridor on the left and right side of the trajectory provides an additional meter of distance to the surroundings. This serves as an orientation aid, for example for parked vehicles whose doors could be opened.

Furthermore, the HMI displays the lateral and longitudinal acceleration and deceleration g-force values based on the current speed change. This information supports the RD in stability control in dynamic driving situations such as cornering or braking maneuvers. 

\subsection{Remote Driving System Operational Design Domain}
\label{RDS ODD}
The definition of the ODD ensures that the RDS guarantees the required connectivity and controllability within a defined environment \cite{hans2024backedautonomy}. The ODD of the RDS is specified with regard to the environment, traffic structure, speed limits, weather conditions and other relevant parameters. The RDS used for this study operates exclusively in an urban environment, specifically in Las Vegas, Nevada. The urban environment offers a variety of traffic and infrastructure conditions that the RDS has to cope with. However, the ODD excludes specific parameters such as:
\begin{itemize}
    \item \textbf{Speed limitation:} Streets in the defined system ODD are limited to a maximum of \mbox{35 mph}. 
    \item \textbf{Weather conditions:} Specific conditions such as snow, ice and rain are excluded from the ODD in this study.
    \item \textbf{Time of day:} The use of RDS is limited to driving during daylight hours in this study.
    \item \textbf{Sufficient and stable connectivity:} As \mbox{Hans et al. \cite{hans2023operational}} point out in their methodology for the ODD definition, connectivity is an essential prerequisite for the operation of the RDS and must be sufficiently guaranteed for the reliable exchange of information of the driving environment. A stable communication link is a key ODD requirement for the safe operation of the RDS, which has been identified as one of the main limitations in previous studies \cite{hans2024backedautonomy, Kang2018Augmenting}.
    \item \textbf{Human in the loop:} In contrast to ADS applications, the RDS does not require the system itself to take over the driving task completely, but instead relies on a human-in-the-loop approach. The task of the RD therefore requires specific qualifications that focus primarily on the control of the vehicle in the defined ODD.
\end{itemize}

\subsection{Remote Driver Training}
\label{RDTraining}
To ensure the controlled and effective use of an RDS, special education and training of the RDs is required. Even though the RDs considered in this work already have a driver's license for at least two years, this is not sufficient to meet the requirements of remotely controlling a vehicle due to the inherent limitations of RDSs as described in \mbox{Section \ref{remotedriving}}. Scientific research emphasizes that remote control of the system and response to vehicle-specific control requirements require additional skills from experienced \mbox{drivers \cite{hans2025evaluationremotedriverperformance, Schwindt-Drews2024}}. The training program of the RDs considered in this study is described in detail \mbox{by Hans et al. \cite{Hans2023Academy}} and aims to equip RDs with specific knowledge for the remote operation. 

\begin{table*}[b]
\centering
\renewcommand{\arraystretch}{1.2} 
\begin{tabular}{p{1.5cm}p{2.6cm}p{2.3cm}p{2.3cm}p{1.8cm}p{1.8cm}p{1.5cm}}
\hline
\textbf{RD level} & \textbf{Cumulative driving \mbox{Experience} [km]} & \textbf{Remotely driven \mbox{distance [km]}} & \textbf{Remotely driven \mbox{duration [mins]}} & \textbf{Average speed [mph]} & \textbf{Average speed [km/h]} & \textbf{RD contributed}\\ \hline \hline
RD--L1 & \(<\)200    & 4,132.56 & 13,202.53 & 11.16 & 17.96 & 25\\ \hline
RD--L2 & 200--500  & 5,494.06 & 14,141.65 & 14.03 & 22.58 & 18\\ \hline
RD--L3 & 500--800  & 4,665.04 & 11,510.48 & 14.35 & 23.09 & 16\\ \hline \hline  
All levels & 0--800 & 14,291.65 & 38,854.67 & 13.02 & 20.96 & 25\\ \hline 
\end{tabular}
\caption{Overview of remotely driven distance, remotely driven duration, and average speed across Remote Driver (RD) experience levels.}
\label{overview_td_level}
\end{table*}

\section{Data Preparation and Filtering}
\label{AnalysisDataPreparation}
The remote driving data set underwent rigorous preparation and filtering to ensure the reliability and validity of the analysis. Therefore, the following criteria were applied: 

\begin{itemize}
    \item \textbf{Analysis period:} The analysis covered the period from \mbox{August 1, 2023}, to \mbox{December 01, 2024}, ensuring sufficient data for trend identification. 
    \item \textbf{Session length exclusion:} Sessions shorter than \mbox{0.1 m} were excluded to avoid incomplete or irrelevant data points due to performed vehicle start-up checks. 
    \item \textbf{System Under Test exclusion:} Data recorded under specific test conditions of the Vay RDS was removed to avoid biases.
    \item \textbf{ODD compliance:} Only data from within the defined ODD, as specified in Section \ref{RDS ODD}, was considered.
    \item \textbf{Public roads only:} Data from restricted testing areas was excluded.
    \item \textbf{Driver roles:} Data was filtered to include only designated RD, as defined in Section \ref{RDTraining}, until a remote driving experience level of \mbox{800 km}.
\end{itemize}

In addition, this study focuses exclusively on disengagements where the SD appropriately assumed control, such as in cases of errors made by the RD. The responsibility for these disengagements lies with the RD, whereby cases attributable to other traffic participant misbehavior or technical failures are excluded. Only disengagements where the RD did not react or did not react appropriately are analyzed. As these are RD training data sessions, a SD was present in all analyzed cases to ensure that control was taken if necessary.

\begin{figure}[ht]
\centerline{\includegraphics[width=3.3in]{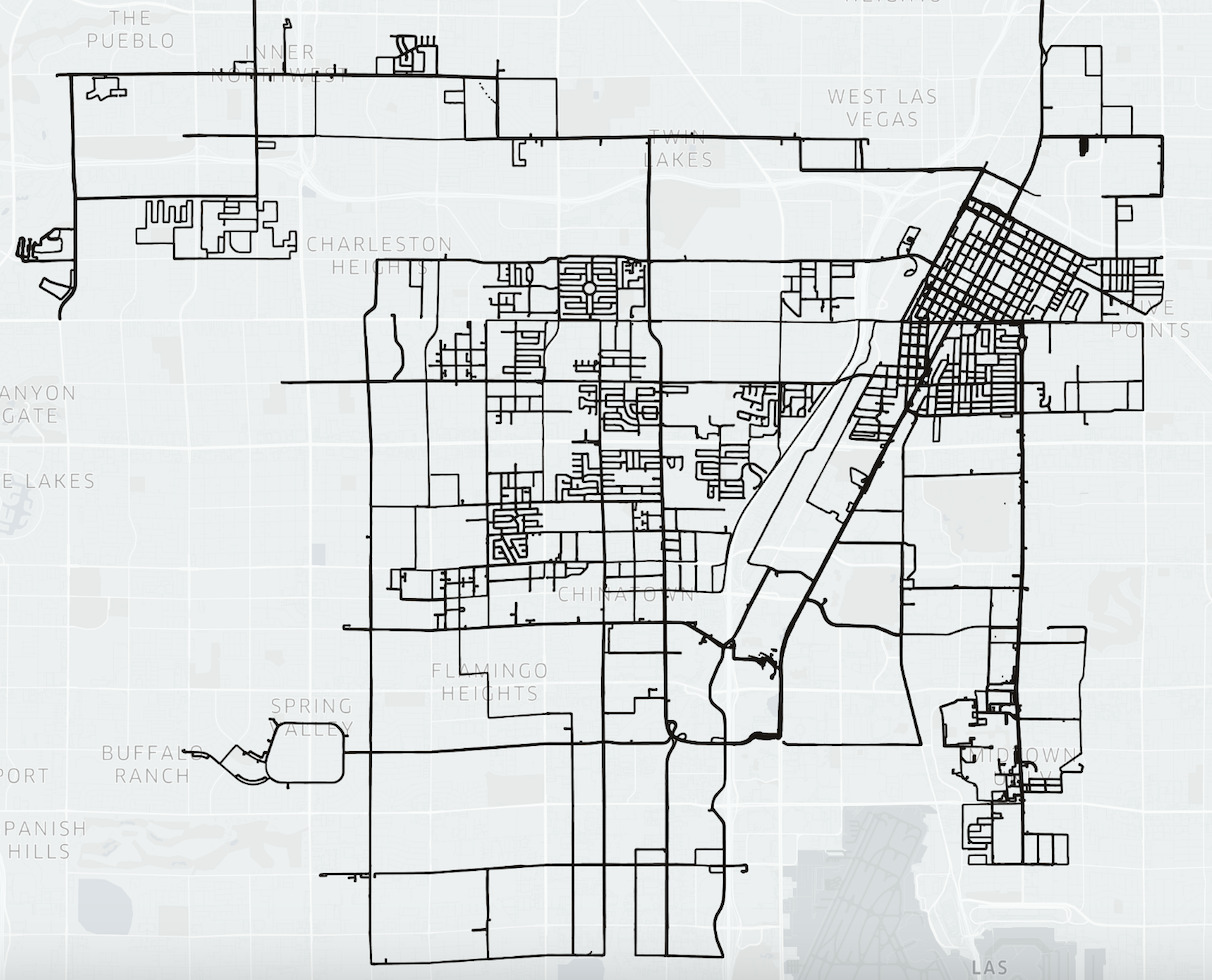}}
\caption{Extract of the geographical part of the Remote Driving System (RDS) Operational Design Domain (ODD) and distribution of the remotely driven distance of more than \mbox{14,000 km} in Las Vegas, Nevada, US.
\label{Figkepler}}
\end{figure}

\section{Evaluation}
\label{evaluation}
The analysis in this work quantifies the relationship between RD's driving experience and the number of disengagements and its causes using data collected from a group of 25 RDs. These operated in an ODD, which was designed according to the ODD qualification process of Hans et al. \cite{hans2023operational}. The RDs were aged between 22 and \mbox{44 years} (\mbox{$M = 30.44$}, \mbox{$SD = 4.81$}), where 21 identified as men and 4 identified as women. All participants completed the standardized remote driving training program from Section \ref{RDTraining} in advance. This ensures a uniform level of competence so that the influence of cumulative remote driving experience on driving performance could be investigated without differences in basic skills acting as a confounding factor.

The study covers a total driving distance of \mbox{14,291.65 km}, spread over several remote driving sessions within the ODD defined in Section \ref{RDS ODD} and shown in \mbox{Fig. \ref{Figkepler}}. During these sessions, performance metrics were continuously collected to analyze the relationship between cumulative remote driving experience, number of disengagements and their reasons in detail. To systematically assess the impact of various influencing factors on remote driving performance, the data analysis considered the three dimensions (1) the driving speed, (2) the level of driving experience, and (3) the driving scenario type. RDs were first categorized into three experience levels based on their cumulative remotely driven distance. Table~\ref{overview_td_level} provides an overview of the total remotely driven distance and driving duration for each experience level, as well as aggregated values across all levels. In addition to this, each driving event was classified by scenario type (e.g., intersection vs. straight road segments) and by speed range (slow, medium, or high), based on predefined thresholds. This multi-factorial categorization enabled a differentiated analysis of performance-related indicators across varying operational contexts.

\subsection{Evaluation of average driving speed}
The average driving speed per remotely driven session increases as shown in Table~\ref{overview_td_level}, indicating a relationship between experience and operational efficiency. RD--L1 drivers exhibit a lower average speed of 11.16\,mph, compared to 14.03\,mph for RD--L2 and 14.35\,mph for RD--L3 drivers. This difference reflects growing driving competence and possibly greater trust either in the RDS or in one’s own abilities as experience increases. This is confirmed by a statistically significant, moderately positive correlation between certification level and average speed. Specifically, a Pearson correlation of $r = 0.446$ ($p < 0.001$) and a Spearman correlation of $\rho = 0.452$ ($p < 0.001$) suggest that more experienced RDs tend to operate at higher average speeds. At the same time, the reduced speed observed that RD--L1 may reflect a more cautious driving style typical of the early stages of driving experience. The small difference in average speed per session between RD--L2 and RD--L3 suggests that the most substantial gain in driving competence occurs during the transition from RD--L1 to RD--L2, while further experience beyond this point has a diminishing impact on average speed.

\subsection{Number of Disengagements over Driving Experience}
\label{disengagementRDexperience}
\begin{figure}[ht]
\centerline{\includegraphics[width=3.5in]{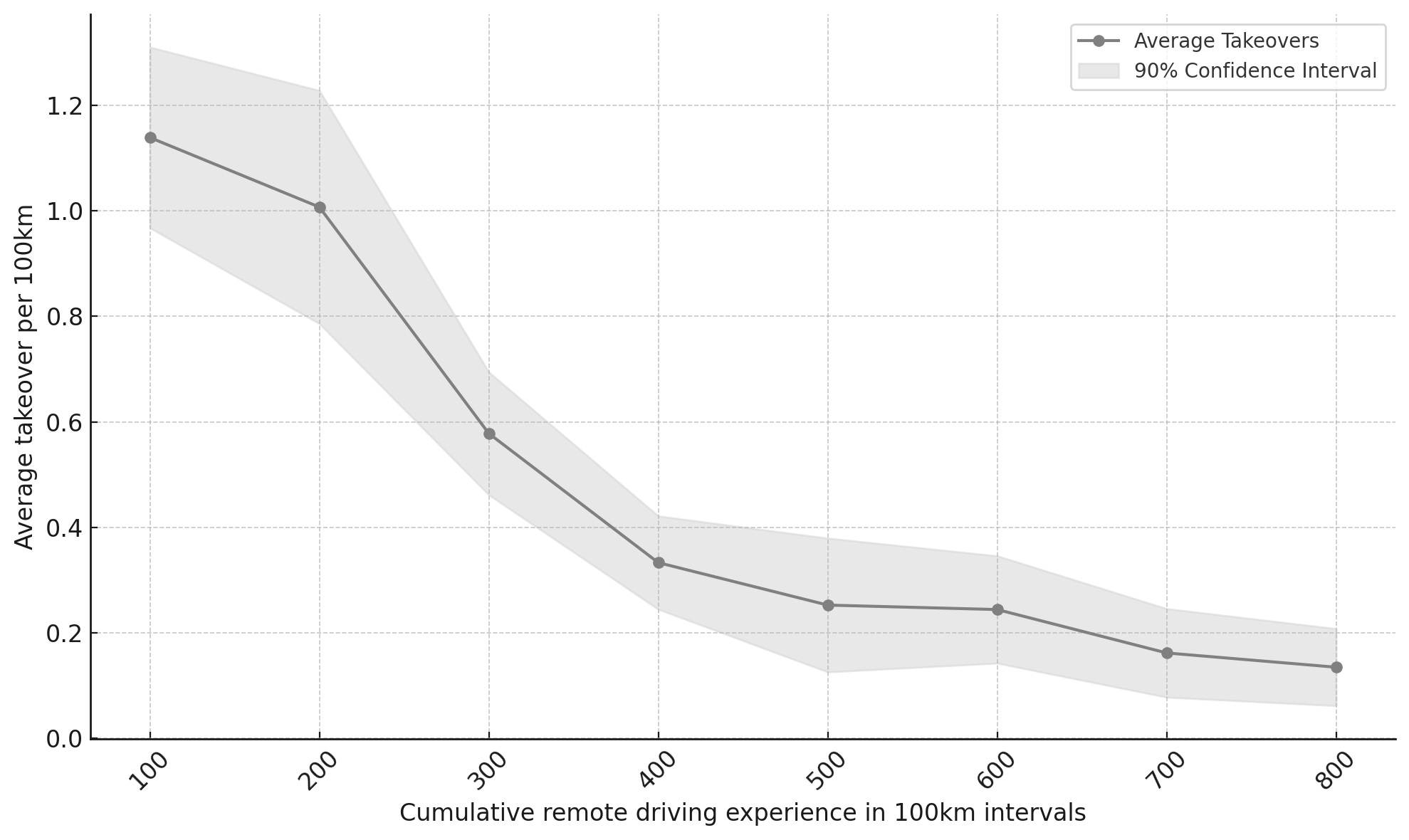}}
\caption{Evolution of the count of Safety Driver (SD) intervention over remote driving experience.
\label{FigdisengagementRDexp}}
\end{figure}

With regards to the number of disengagements depending on the remote driving experience, Fig. \ref{FigdisengagementRDexp} shows the correlation between the average number of SD interventions per \mbox{100 km} and the cumulative RD experience in intervals of \mbox{100 km}. 

The curve shows a clear decreasing trend in the number of interventions per \mbox{100 km} with increasing cumulative remote driving experience. The data suggests that there is a pronounced learning curve with the RDs. Especially in the first \mbox{400 km} of driving there is a strong reduction in SD interventions. This could be due to a rapid learning of skills or an improved understanding of the RDS. From about \mbox{400--500 km}, there is a plateau in performance improvement. Additional experience reduces interventions only slightly. This indicates that basic skills are developed at an early stage. The confidence interval of 90\% is relatively wide at lower experience levels, indicating a greater spread in the data. This indicates the need for individualized training strategies to account for different starting levels. With increasing experience, the confidence interval becomes smaller, suggesting a more stable and consistent performance over the group of RDs.

\subsection{Disengagement Classification}
\label{disengagement_classification}
The various categories of disengagements in relation to human RD performance are described below. The classification is carried out post-hoc on the basis of video material recorded during the remote drives. The situations in which the SD had to intervene in order to avoid an unsafe or irregular driving situation were analyzed. The classification into these categories creates a basis for the quantitative and qualitative assessment of RD-related disengagements. These are used for the detailed analysis of RD behavior and the identification of specific scenarios that can lead to potential safety-critical situations.

\begin{itemize}
    \item \textbf{Braked too late for signs:} The RD did not slow down the vehicle in time in front of a traffic sign.
    \item \textbf{Traffic light went red:} The RD did not stop the vehicle in time while the traffic light changed from green to red.
    \item \textbf{Impatient for 3rd parties, obstacles:} The RD did not react in time to respond to other traffic participants or objects.
    \item \textbf{Leaving the lane to the left:} The vehicle deviated from the intended lane to the left due to the RD's steering.
    \item \textbf{Leaving the lane to the right:} The vehicle deviated from the intended lane to the right due to the RD's steering.
    \item \textbf{Other:} This category includes other reasons for SD interventions due to the driving performance of the RD and do not fit into the above categories.
\end{itemize}

\subsection{Distribution of Disengagement Reasons}
\label{disengagementInvestigation}
\begin{figure}[ht]
\centerline{\includegraphics[width=3.5in]{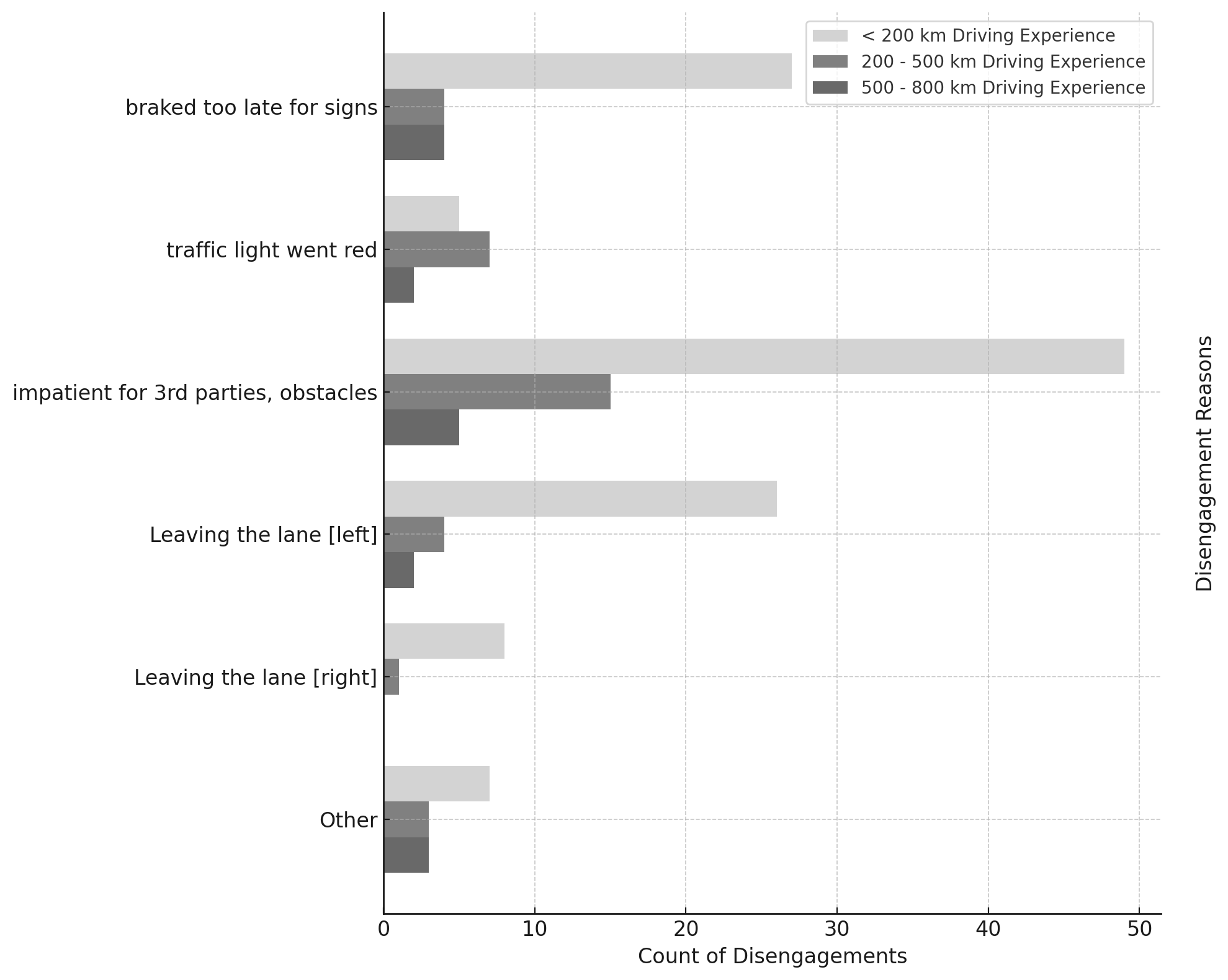}}
\caption{Count per disengagement reasons for different remote driving experience levels.
\label{Figdisengagementreasons}}
\end{figure}

The analysis of the reasons for in total \mbox{183 disengagements} as a function of the RD's cumulative driving experience provides insights into the causes and frequency of safety-relevant situations during remote driving. A detailed analysis of disengagement reasons is done in Chapter VI and is shown in \mbox{Fig. \ref{Figdisengagementreasons}}.

The results of the Chi--Square Tests confirm the findings described above. \mbox{RD--L1} show significantly more frequent SD interventions in almost all categories compared to \mbox{RD--L2} and \mbox{RD--L3}. This is clearly shown in the results of the Chi-Square tests: A \(\chi^2\)--value of 21.70 (\(p = 0.006\)) was found between \mbox{RD--L1} and \mbox{RD--L2}, while the comparison between \mbox{RD--L1} and RD--L3 also showed significant differences with a \(\chi^2\)--value of 19.61 (\(p = 0.012\)). These results illustrate the significant learning progress associated with increasing remote driving experience. In contrast, the results of the \(\chi^2\)--test between \mbox{RD--L2} and \mbox{RD--L3} show no significant differences (\(\chi^2 = 5.70\), \(p = 0.681\)). This confirms that most of the improvement takes place in the first \mbox{500 km} of driving experience and that progress weakens thereafter.

\section{Analysis of Disengagement Reasons}
\label{analysis_dis_reasons}
In this Section, the reasons listed in \mbox{Fig. \ref{Figdisengagementreasons}} of in total \mbox{183 SD interventions} are analyzed in detail for the causes in order to identify possible reason accumulations or patterns.

\subsection{Reason: Braked too late for signs}
\label{brakedtoolateforsigns}
In total 19.13\% of the \mbox{183 disengagements} were recorded in which the RD failed to slow down the vehicle in time before a traffic sign. The frequency of these disengagements decreases with increasing remote driving experience, as shown in \mbox{Fig. \ref{Figdisengagementreasons}}, with 77.14\% of disengagements attributable to RDs with less than \mbox{200 km} of driving experience. 

For \mbox{RD--L1}, 92.59\% of theses interventions consisted of situations where the RD recognized the traffic sign (51.85\% for traffic lights and 48.15\% for stop signs) and verbally called them out, but the braking response was inadequate. This indicates limited adaptation to the long-distance driving environment, possibly due to the partly replaced haptic feedback by visual components, which requires a higher cognitive load and longer adaptation time. This adaptation is more pronounced in experienced RD. In the remaining 7.31\% SD interventions of \mbox{RD--L1}, no reaction was observed as the RD was focused on other road users instead of the road sign. Such behavior was no longer observed for \mbox{RD--L2} and \mbox{RD--L3}, underlining the positive effect of experience on task prioritization and situational awareness. These differences underline the greater decision-making and perception ability of experienced drivers.

\subsection{Reason: Traffic Light went red}
SD interventions in which a traffic light changed to red while the remotely driven vehicle was approaching accounted for 7.65\% of the total number of disengagements analyzed in this study. The number of these incidents remains relatively low across the different experience levels. In this scenarios the speed played a central role, as 14.3\% incidents occurred at speeds of \(<\)12 mph, 21.4\% incidents occurred at speeds of \mbox{12--$<$19 mph} and 64.3\% incidents occurred at speeds of $\geq19$ mph. The error rate at higher speeds shows that these situations are particularly challenging, as the short reaction time available and the high speed make decision-making much more difficult. In 71.4\% of cases, the reasons for these disengagements are based on the fact that the RD made the wrong decision in the respective situation, either because the RD wanted to cross the traffic light or because the RD applied the brakes too hard. In the other 28.6\% of cases, the RD responded too late, too slowly or not at all.

\begin{figure*}[b] 
    \centering
    \begin{subfigure}[t]{0.451\textwidth}
        \includegraphics[width=\linewidth]{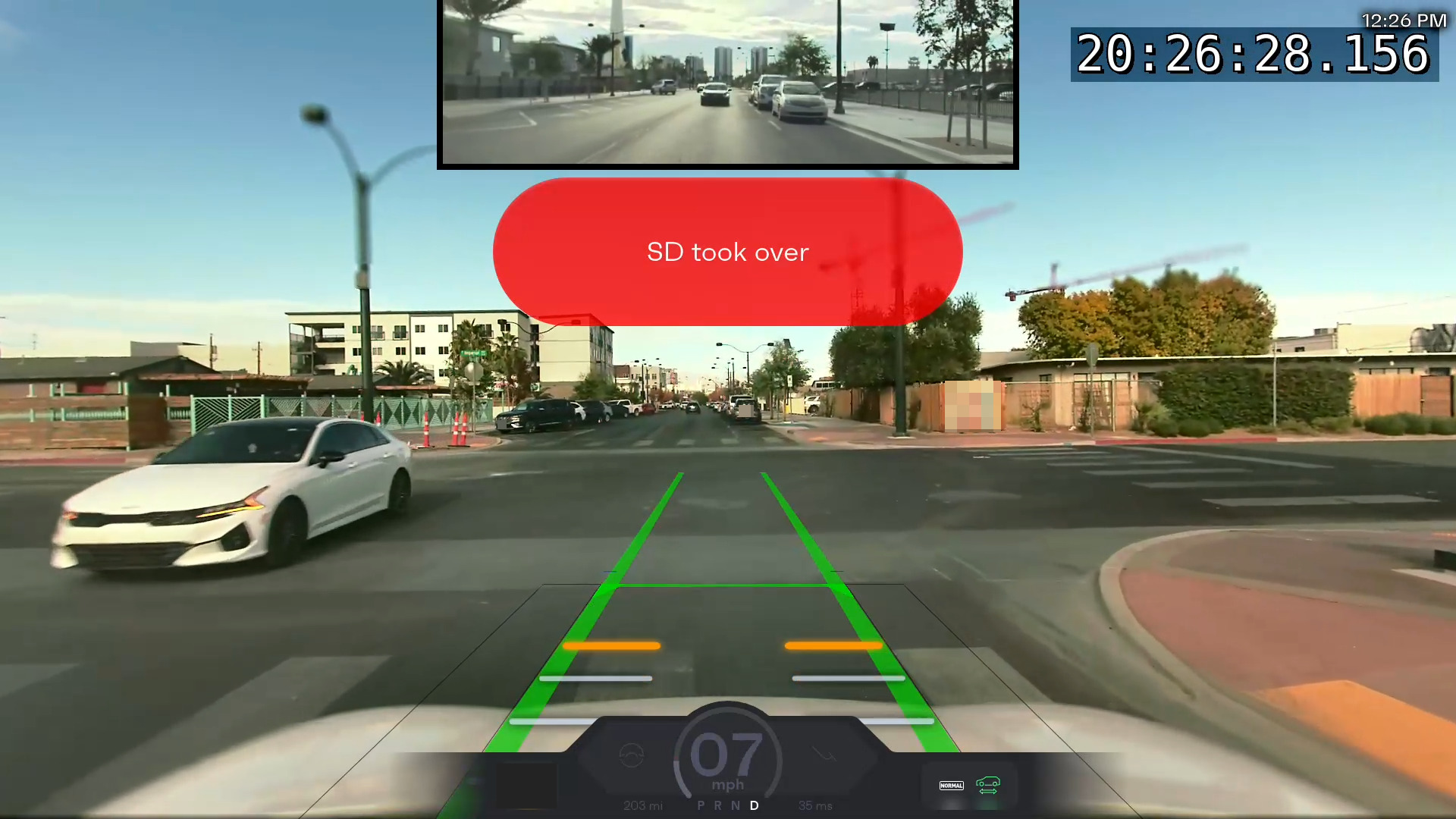}
        \caption{RD braked too late for signs at a speed of $>$6 mph where the RD reacts but does not manage to stop in time in front of the sign.}
        \label{Braked_too_late_for_signs}
    \end{subfigure}
    \hfill
    \vspace{3mm}
    \begin{subfigure}[t]{0.451\textwidth}
        \includegraphics[width=\linewidth]{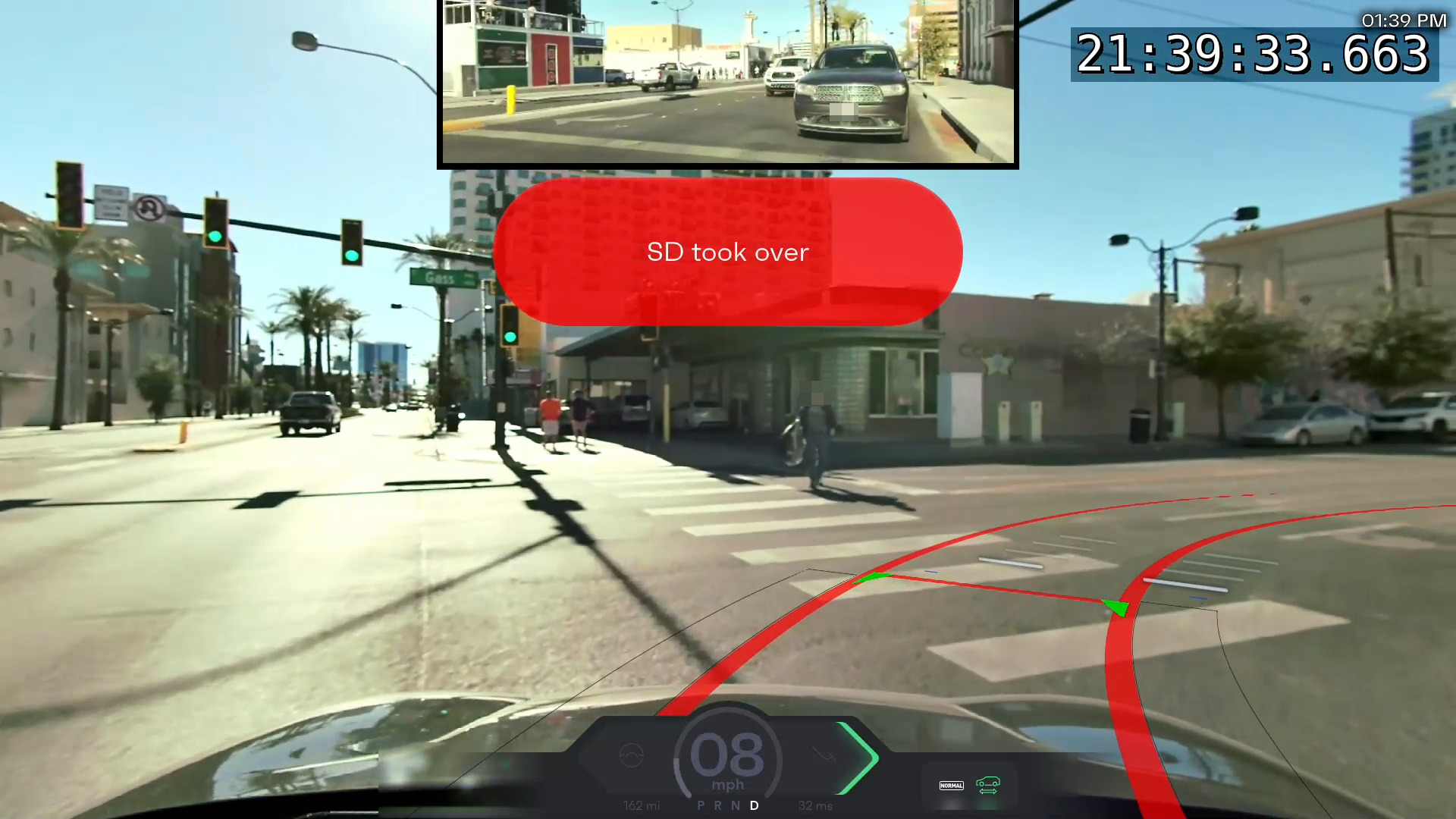}
        \caption{RD is impatient for other traffic participants or obstacles during remote driving at $<$12 mph in the area of an intersection or junction.}
        \label{Fig. Impatient}
    \end{subfigure}
    \hfill
    \begin{subfigure}[t]{0.451\textwidth}
        \includegraphics[width=\linewidth]{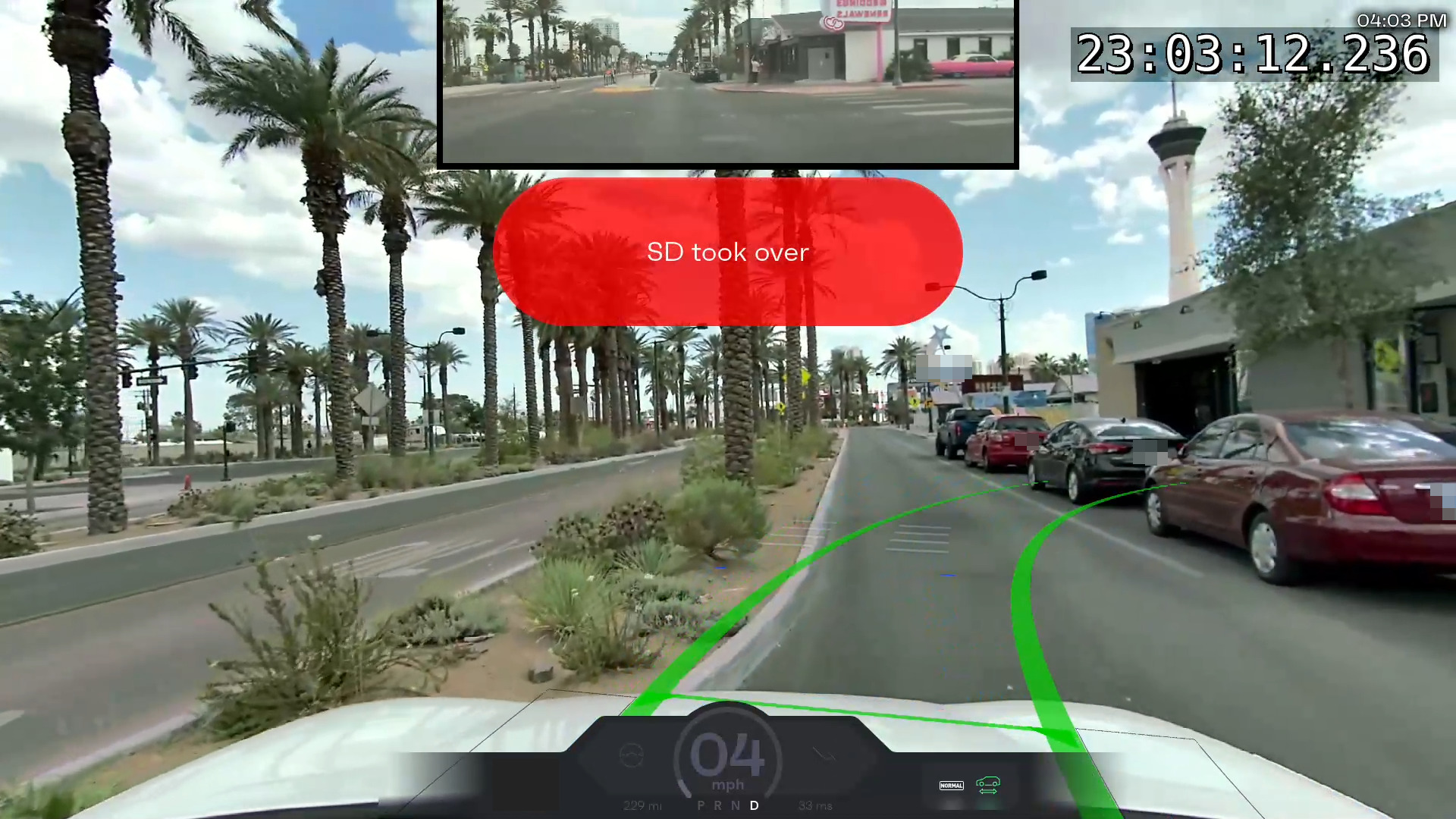}
        \caption{RD left the lane to the left or right at a speed of $<$12 mph.}
        \label{Leaving_lane}
    \end{subfigure}
    \hfill
    \begin{subfigure}[t]{0.451\textwidth}
        \includegraphics[width=\linewidth]{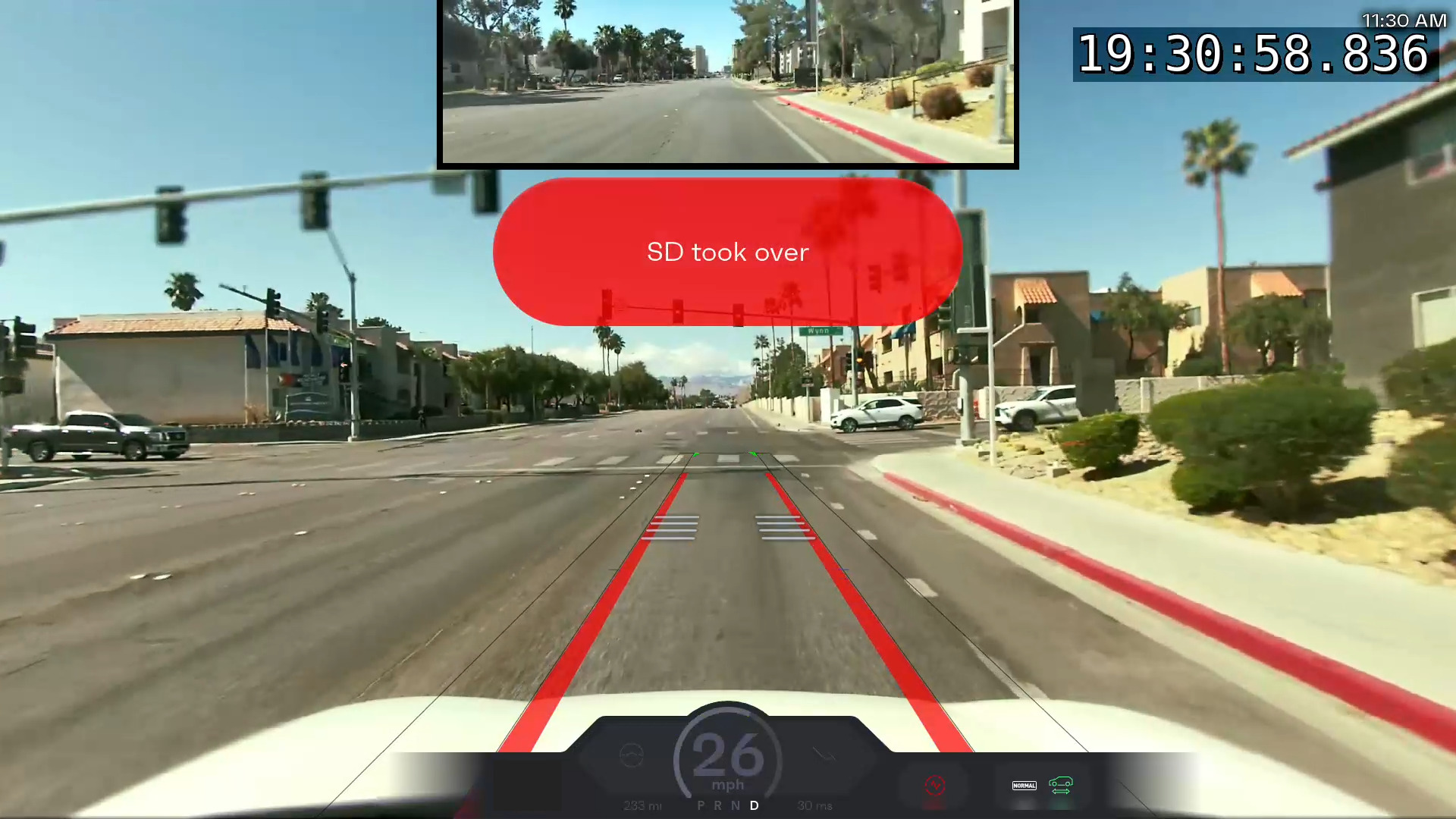}
        \caption{The traffic light changes to red at a speed of $\geq12$ mph where the RD reacts, but in the estimation of the SD makes the wrong decision and the SD takes over.}
        \label{traffic_light_turned_red}
    \end{subfigure}
    \caption{Frequently occurring driving scenarios that lead to the Safety Driver (SD) taking over control during remote driving.}
    \label{FigScenarios}
\end{figure*}

\subsection{Reason: Leaving the lane to the left or right}
Of the total of 183 cases analyzed, 41 can be attributed to specific incidents in which the RD would have left the lane, which corresponds to a share of 22.40\%. In contrast, 78.05\% of these cases are attributable to SD interventions in which the RD would have left the lane to the left. This distribution clearly shows that leaving the lane to the left occurs much more frequently than to the right.

The analysis of performance in terms of driving experience in remote driving scenarios shows clear differences. Of the 41 scenarios analyzed, 82.9\% were performed by \mbox{RD--L1}. While this experience level showed no inappropriate reactions in 35.3\% of cases, there is only one case of \mbox{RD--L2} in comparison, which indicates a higher competence, experience and training for more experienced RDs.

The susceptibility of the RD to errors at low speeds \mbox{(\(<\)6 mph)} is particularly striking, accounting for 43.9\% of the scenarios. If speeds of up to 12 mph are taken into account, this figure rises to 72.7\% of the cases. In these situations, which often occur at intersections or during turning maneuvers, the RDs failed more frequently to make adequate lane following maneuvers. 

\subsection{Reason: Impatient for other traffic participants, obstacles}
Of the disengagements analyzed, a total of 37.7 \% can be attributed to the scenarios where the RD was impatient for other traffic participants or obstacles, as a reason for the SD to take over the control of the vehicle.

The analysis of the performance of RD experience levels in remote driving scenarios shows major differences in the responsiveness and decision-making of these groups. Overall, 71.01\% of these scenarios were performed by \mbox{RD--L1}, while experienced \mbox{RD--L2} were only involved in 21.74\% and \mbox{RD--L3} 7.25\% of cases. The scenarios analyzed covered different geometries and speeds. Intersections and junctions were the most common type of driving situation with 66.67\%, followed by straight roads with 23.19\%. 

Another important factor is speed, as low speeds \mbox{(\(<\)6 mph)} were the most error-prone with 60.9\% of errors, while medium speeds (6--$<$19 mph) accounted for 22.7\% and high speeds ($\geq19$ mph) only 17.4\% of errors. This indicates that low speeds, which are often associated with more complex scenarios such as intersections, present a particular challenge, as visualized in the specific scenario of \mbox{Fig. \ref{Fig. Impatient}}.

\section{Limitations}
\label{Limitation}
This study has several limitations that should be considered when interpreting the results:  

First, the results are specific to the RDS and its ODD defined for this study based on a specific use case. While they provide valuable insights, they may not fully translate to other vehicle types such as trucks or other ODDs such as desert environments without further validation. Nevertheless, these results can serve as a starting point for similar applications in other ODDs or vehicle types and provide a basis for further research.  

Secondly, the analysis does not take into account individual RD parameters such as situational awareness or workload. By neglecting these individual parameters, the study potentially misses important insights into the specific challenges and complexities that RDs face in real-world scenarios.

Finally, the necessity of the SD interference as a precautionary measure cannot be clearly answered. Although there is an mistake by the RD, whether this actually leads to safety-critical scenarios is not answered in this work.

\section{Discussion}
\label{conclusion}
The analysis focuses on the disengagements by the RD, which are primarily due to human perception and decision-making errors, while in the case of ADS they result primarily from technical limitations, especially in the area of environment recognition and decision logic under \mbox{uncertainty \cite{cummings2024identifying}}. While RD can demonstrably reduce its susceptibility to errors through targeted training, ADS requires structural improvements to AI architectures and more robust behavior in complex traffic scenarios. The results make it clear that the causes of disengagements depend heavily on the system type and that differentiated optimization strategies are required accordingly. 

The analysis of 183 disengagements caused by RD driving mistakes provides insights for the further development of RDSs. The results show that cumulative remote driving experience is a decisive factor in reducing disengagements. In particular, the decline in safety-related SD interventions is achieved within the first \mbox{400 km} driven, indicating a pronounced learning curve. After this phase, there is a plateau, which shows that the basic skills of the RD are developed at an early stage of their driving experience in addition to their previous training. This finding is supported by the statistical analysis, which shows significant differences in the reasons for disengagement between RDs with less than \mbox{200 km} experience and RDs with \mbox{200--800 km} experience. In this analysis, four specific scenarios, shown in \mbox{Fig. \ref{FigScenarios}}, were identified.

The first scenario, shown in \mbox{Fig. \ref{Braked_too_late_for_signs}}, at a speed of $\geq6$ mph where the RD reacts but does not manage to stop in time in front of a sign. This maneuver accounts for 18.03\% of all disengagements, of which 13.66\% can be attributed to RDs with a driving experience of \mbox{\(<\)200 km}. This means that this event occurs every \mbox{165.30 km} for the RD-L1 and only every \mbox{1,269.89 km} for the \mbox{RD--L2} and \mbox{RD--L3}. 

A total of 23.50\% of the disengagements are due to the scenario in which the RD is impatient for other traffic participants or obstacles during remote driving at \(<\)12 mph in the area of an intersection or junction (see \mbox{Fig. \ref{Fig. Impatient}}). 

Leaving the lane during turning left or right at a speed of \(<\)12 mph is responsible for a total of 12.57\% of the disengagements and represents scenario 3, which is shown in \mbox{Fig. \ref{Leaving_lane}}. Here, the RD-L1 are responsible for a full 11.48\%, which means that the \mbox{RD--L1} experience this scenario every 196.79km, while RD with a remote driving experience of \mbox{200--800 km} only experience this scenario every \mbox{5,079.55 km}. The reason for this is possibly the additional latency, the insufficient haptic feedback or distorted spatial perception.

The proportion of the scenario, visualized in \mbox{Fig. \ref{traffic_light_turned_red}}, where a traffic light changes to yellow/red at a speed of $\geq12$ mph is 6.01\%. Here the RD reacted, but according to the assessment of the SD for wrong and the SD had to take over based on the respective situation.

The results underline the need to focus training programs on the initial learning phase in order to effectively promote the development of basic driving skills. This can significantly increases controllability in remote driving operations, especially in demanding urban traffic environments.

Future work should investigate whether the identified problems and specific scenarios described in this work can already be assessed with other driving metrics, so that constructive or instructive measures can be taken at an early stage to prevent SD interventions. In addition, possible measures to improve training should be discussed, such as the targeted use of simulator training to systematically prepare RDs for typical error scenarios and improve their situational decision-making behavior. At the same time, it remains unclear whether the RD errors analyzed in the study would actually have led to safety-critical situations if no intervention had taken place. Although the categorization of disengagement is based on observable driving errors, a clear causal relationship between RD behavior and real hazard potential is not conclusively established. Future research should therefore differentiate more strongly between actual and potential hazards.
\bibliography{Hans_biblio}

\end{document}